\documentclass[%
 amsmath,amssymb,
 aps, physrev,
]{revtex4-2}
\usepackage{graphicx}% Include figure files
\usepackage{dcolumn}% Align table columns on decimal point
\usepackage{bm}% bold math
\usepackage{braket}
\usepackage[table,xcdraw]{xcolor}

\begin{document}

%\preprint{APS/123-QED}

\title{\textbf{About Matter Effects on Neutrino Phases in Mixing Matrix} 
}% 

% \altaffiliation[Also at ]{Physics Department, XYZ University.}%Lines break automatically or can be forced with \\
\author{Holanda, P. C. de }
 \email{Contact author: holanda@unicamp.br}
\author{Jales, M.}
\author{Walter, E.}
 \affiliation{%
Instituto de F\'\i sica Gleb Wataghin - IFGW \\
Universidade Estadual de Campinas (UNICAMP) 
%  Include all institutions where the work was conducted: department or division, institution, city, state (if relevant), and country, in this order.
}%

%\title{\textbf{Neutrino phase shift in matter} }
%
%
%\author{}
%\author{}
%% \email{Contact author: holanda@unicamp.br}
%\author{e}
%\affiliation{%
%
\date{\today}

\begin{abstract}
  As is widely known, the flavour and mass eigenbasis in neutrino sector can be related by a reduced unitary matrix, described by 3 real parameters and one phase, for Dirac neutrinos, and 2 extra phases if neutrinos are Majorana particles. Neutrino flavour oscillation experiments are insensitive to Majorana phases, which leads to using the PMNS matrix with only one phase to describe flavour transitions. In this paper, we point out one feature of such choice that is commonly ignored: when neutrinos propagate in matter, the reduced unitary matrix PMNS does not properly diagonalize the neutrino system. We present some numerical calculations and analytical estimations to illustrate such statement, and discuss some scenarios where this effect can be probed.
\end{abstract}

\maketitle

\newpage
\section{Introduction}

Neutrino physics has stood out in the field of elementary particles by producing a remarkably rich phenomenology. One of the key aspects of this phenomenology lies in the fact that weak interactions via charged currents are defined in a basis different from the mass basis, which are related through a unitary transformation.

Different experiments are sensitive to different combinations of the entries of this unitary matrix. Notable examples include neutrinoless double beta decay experiments~\cite{Pascoli:2005zb}, the high energy tail of the beta decay spectrum~\cite{KATRIN:2021uub}, and flavor oscillations~\cite{Gonzalez-Garcia:2007dlo}. A complete understanding of the structure of the mixing matrix requires a combined analysis of these various experimental results~\cite{Capozzi:2021fjo}.

In particular, the six phases present in a generic 3×3 unitary mixing matrix can lead to a variety of observable effects. Three of these phases can be  considered unphysical, as they can be absorbed into the definition of the neutrino eigenstates and leave no experimental signatures. Of the three remaining phases, two appear only if neutrinos are Majorana particles. These Majorana phases can affect neutrinoless double beta decay experiments but do not influence flavor oscillation dynamics - at least when effects from physics beyond the Standard Model are not present (see for instance~\cite{Pradhan:2024qrp} and~\cite{Benatti:2001fa} for some counterexamples). %~\cite{Parveen:2024cff} 
The final phase is associated with CP symmetry violation in the neutrino sector.

When neutrinos propagate through matter, additional effects must be taken into account due to forward scattering with electrons~\cite{Wolfenstein:1977ue,Mikheyev:1985zog}. These matter effects can be described in terms of effective values for the entries of the mixing matrix in matter~\cite{Cherchiglia:2024bhl}. In this article, we show that the argument made in the previous paragraph must be revisited in this context. In particular, it is no longer possible to absorb the unphysical phases through a single redefinition of the lepton fields.

This paper is organized as follows. On sec.~\ref{sec:nuphases} we introduce a brief review of the phases on neutrino mixing angles and present the issue we approach, numerically, on sec.~\ref{sec:numerically}. Some analytical considerations are presented on sec.~\ref{sec:analytical}. We conclude on sec.~\ref{sec:conclusions}.

\section{Neutrino phases}\label{sec:nuphases}

Neutrinos do not propagate in the same eigenspace in which they interact, which leads to the known mechanism of mass-induced flavour oscillation. The conversion between mass and flavour eigenbasis is implemented through a $3\times 3$ unitary neutrino mixing matrix $U$:
\[
%\left(\begin{array}{c}
%\nu_e \\ \nu_\mu \\ \nu_\tau 
%\end{array}\right)=
%U
%\left(\begin{array}{c}
%\nu_1 \\ \nu_2 \\ \nu_3 
%\end{array}\right)
\nu_\lambda=U_{\lambda i}\,\nu_i
\]
where $\nu_\lambda$ ($\lambda=e,\mu,\tau)$ form the neutrino interacting basis and $\nu_i$ $(i=1,2,3)$ are the neutrinos with well-defined mass.

The most general unitary matrix in 3-dimensions involves 3 real mixing angles and 6 phases. Still, not all these angles have physical meaning or are independent parameters. %, and some of those extra phases can be absorbed in the definition of lepton fields.
Following the notation presented in sec.~6 of ~\cite{Bilenky:2018hbz}, a unitary matrix can be written as:
\[
%{\rm diag}(e^{i\beta_1},e^{i\beta_2},e^{i\beta_3})
u_{li}=e^{i\beta_l}U_{il}^De^{i\alpha_l}
\]
where $i,l$ runs from 1 to $N=3$, $N$ being the dimension of the unitary matrix, and $U^D$ is the mixing matrix for Dirac neutrinos, which contains one phase (see~\cite{Bilenky:2018hbz} for details). We can set $\alpha_1=0$ without loss of generality. Besides, given the structure of the leptonic charged current, the phases $\beta_l$ can be absorbed in the definition of the charged lepton field, resulting in a mixing matrix with only 3 phases:
\begin{equation}
U=U_{\text{PMNS}}U_{maj}
\label{eq:Umatrix}
\end{equation}
where the PMNS matrix involves the three mixing angles and the CP phase:
\begin{eqnarray}    
U_{\text{PMNS}}&=&\left(
\begin{array}{ccc}
    1 & 0       & 0 \\
    0 &  c_{23} & s_{23} \\
    0 & -s_{23} & c_{23} 
\end{array}
\right)
\left(
\begin{array}{ccc}
    c_{13}  & 0 & s_{13}e^{-i\delta_{CP}} \\
    0       & 1 & 0 \\
    -s_{13}e^{i\delta_{CP}} & 0 & c_{23} 
\end{array}
\right)
\left(
\begin{array}{ccc}
    c_{12}  & s_{12}  & 0    \\
    -s_{12} & c_{12} & 0 \\
    0       & 0 & 1 \\
\end{array}
\right)\nonumber\\
&=&
\left(
\begin{array}{ccc}
    c_{12}c_{13} & s_{12}c_{13} & s_{13}e^{-i\delta_{CP}} \\
    -s_{12}c_{23}-c_{12}s_{23}s_{13}e^{+i\delta_{CP}} &  c_{12}c_{23}-s_{12}s_{23}s_{13}e^{+i\delta_{CP}} & s_{23}c_{13} \\
    s_{12}s_{23}-c_{12}c_{23}s_{13}e^{+i\delta_{CP}} & -c_{12}s_{23}-s_{12}c_{23}s_{13}e^{+i\delta_{CP}} & c_{23}c_{13} 
\end{array}
\right)
\label{eq:PMNS}
\end{eqnarray}
and $U_{maj}$ contains the Majorana Phases:
\begin{equation}
U_{maj}=\left(
\begin{array}{ccc}
    1 & 0       & 0 \\
    0 &  e^{i\alpha_2} & 0 \\
    0 & 0 & e^{i\alpha_3} 
\end{array}
\right)
\label{eq:majoranaphases}
\end{equation}

\subsection{Matter Effects}

Besides the neutrino masses and mixing, matter effects can also interfere with the neutrino evolution.
To diagonalize the $3\times 3$ neutrino Hamiltonian in matter we have to find
the mixing matrix $\tilde U$:
\begin{equation}
\tilde{U}^\dagger\left[U\lambda U^\dagger+V\right]\tilde{U}=\tilde\lambda
\label{eq:mixinmatter}
\end{equation}
where $\lambda=diag(\lambda_1,\lambda_2,\lambda_3)$ is the matrix containing the eigenvalues in vacuum, $V=diag(V_{CC},0,0)$ describes the matter potential, and the tilde denotes the effective values of the referring quantities in matter.

In principle, to diagonalize the Hamiltonian in matter, we should turn back to the most general unitary matrix and review the arguments for absorbing phases. It is clear that, if the mixing matrix $U$ is real, then $\tilde{U}$ is also real. 
\begin{align}
  U_{PMNS} &\rightarrow \tilde{U}_{PMNS}
\end{align}

However, if $U$ is written as in eq.~(\ref{eq:Umatrix}), we could think, wrongly, that $\tilde U$ also has the same format. We show here that this is not the case. 

\begin{align}
  U_{\text{PMNS}}U_{maj} &\not\rightarrow \tilde{U}_{\text{PMNS}}\tilde{U}_{maj}
\end{align}

Starting with the Majorana phases, any mixing matrix that diagonalizes the Hamiltonian in matter has the following freedom:
\begin{equation}
\tilde{U} \rightarrow \tilde{U}\left(
\begin{array}{ccc}
    e^{i\gamma_1} & 0       & 0 \\
    0 &  e^{i\gamma_2} & 0 \\
    0 & 0 & e^{i\gamma_3} 
\end{array}
\right)
\label{eq:phasefreedom}
\end{equation}
as can be trivially satisfied by inserting eq.(\ref{eq:phasefreedom}) in eq.(\ref{eq:mixinmatter}). 
So, at least in terms of the neutrino mixing matrix, the Majorana phases are not affected by matter effects. Thereby we will proceed by looking for a mixing matrix in the form:
\begin{equation}
\tilde{U}=\tilde{U}_{ph}\tilde{U}_{\text{PMNS}}
\label{eq:Uph}
\end{equation}
where $\tilde{U}_{ph}$ accounts for the unphysical phases previously absorbed by the charged lepton fields:
\[
\tilde{U}_{ph}=\left(
\begin{array}{ccc}
    1 & 0       & 0 \\
    0 &  e^{i\alpha} & 0 \\
    0 & 0 & e^{i\beta} 
\end{array}
\right)
\]
We choose $(\tilde{U}_{ph})_{11}=1$ since for oscillation analysis we are only interested in phase differences.

\section{Numerical diagonalization}\label{sec:numerically}

The result of numerically finding $\tilde{U}$  has some degeneracies. For instance, a global phase on any column  (the Majorana phases) can be absorbed in the corresponding mass eigenstate $\nu_i$, analytically stated in eq.(\ref{eq:phasefreedom}). And since these phases have no effect on flavour conversion, it is always possible to rewrite the mixing matrix in matter as:
\begin{equation}
\tilde{U}=\tilde{U}_{ph}\tilde{U}_{\text{PMNS}}
\label{eq:Uph_matter}
\end{equation}
with the following procedure:

\begin{enumerate}
    \item Numerically diagonalize $H_{mat}$;
    \item Calculate $\tilde{\theta}_{12}$, $\tilde{\theta}_{13}$, and $\tilde{\theta}_{23}$ through absolute values of $\tilde{U}_{11}$, $\tilde{U}_{12}$, $\tilde{U}_{13}$, $\tilde{U}_{23}$, and $\tilde{U}_{33}$;
    \item Use the Jarlskog invariant to calculate $\tilde{\delta}_{CP}$ in the following way:
    \begin{eqnarray*}
    J&=&U_{\mu 3}U_{e2}U_{\mu 2}^*U_{e3}^*=
    \tilde{s}_{12}\tilde{s}_{23}\tilde{s}_{13}\tilde{c}_{13}^2e^{+i\tilde{\delta}_{CP}}(\tilde{c}_{12}\tilde{c}_{23}-\tilde{s}_{12}\tilde{s}_{23}\tilde{s}_{13}e^{-i\tilde{\delta}_{CP}})\\
    &=& \tilde{s}_{12}\tilde{s}_{23}\tilde{s}_{13}\tilde{c}_{13}^2(\tilde{c}_{12}\tilde{c}_{23}\,cos\tilde{\delta}_{CP}-\tilde{s}_{12}\tilde{s}_{23}\tilde{s}_{13})+i\tilde{s}_{12}\tilde{c}_{12}\tilde{s}_{23}\tilde{c}_{23}\tilde{s}_{13}\tilde{c}_{13}^2\sin\tilde{\delta}_{CP}
    \end{eqnarray*}
    \begin{eqnarray*}
    &\rightarrow&~
    \sin\tilde{\delta}_{CP}=\frac{Im\left[J\right]}{\tilde{s}_{12}\tilde{c}_{12}\tilde{s}_{23}\tilde{c}_{23}\tilde{s}_{13}\tilde{c}_{13}^2}\\
    &\rightarrow&~
    \cos\tilde{\delta}_{CP}=\frac{1}{\tilde{c}_{12}\tilde{c}_{23}}\left(\frac{Re\left[J\right]}{\tilde{s}_{12}\tilde{s}_{23}\tilde{c}_{13}^2\tilde{s}_{13}}+\tilde{s}_{12}\tilde{s}_{23}\tilde{s}_{13}\right)
    \end{eqnarray*}
    \item Use the freedom in eq.(\ref{eq:phasefreedom}), absorbing the necessary phase on $\nu_1$, $\nu_2$, and $\nu_3$ to have real values for $\tilde{U}_{e1}$, $\tilde{U}_{e2}$, and an imaginary part of $\tilde{U}_{e3}$, befitting the obtained value of $\tilde{\delta}_{CP}$;
    \item Determine the remaining phases from the complex phase of $\tilde{U}_{\mu 3}$ and $\tilde{U}_{\tau 3}$.
\end{enumerate}  
After that, the full mixing matrix is written as in eq.~(\ref{eq:Uph_matter}). In the following we present our results. If not stated otherwise, in all panels we use the values below for the oscillation parameters:
\begin{gather*}
\Delta m^2_{21}=8\,10^{-5}~\textrm{eV}^2~;~ 
\Delta m^2_{31}=2.5\,10^{-3}~\textrm{eV}^2 \\
\theta_{12}=0.59~;~
\theta_{13} = 0.148~;~
\theta_{23}= 0.738 \\
E_\nu=10\,\textrm{MeV}
\end{gather*}

\begin{figure}[thb]
  \vspace{1.0cm}
  \centering
    \includegraphics[width=150mm]{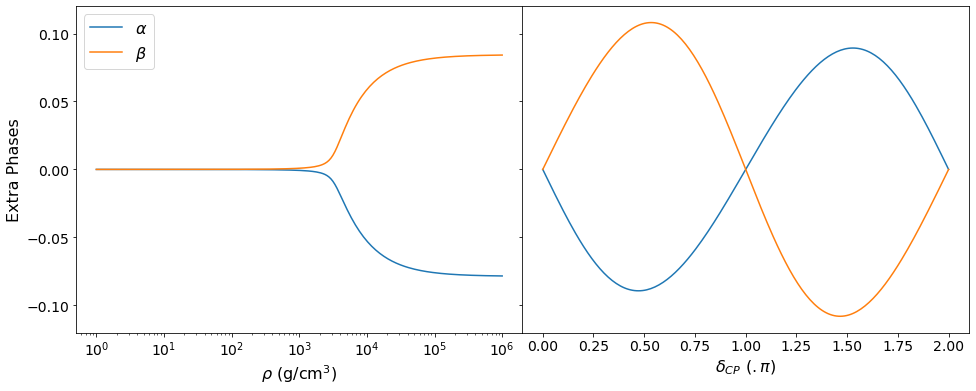} 
    \caption{Value of the new phases as a function of density and CP-phase. In the left panel we set $\delta_{CP}=0.3\pi$, while in the right panel, we set $\rho=10^5$ g/cm$^3$. The other neutrino parameters are described in the text.}
    \label{fig:1}
\end{figure}

In Fig.~\ref{fig:1} we present the value of the new phases as a function of density, keeping $\delta_{CP}=0.5\pi$ fixed (left panel), and as a function of CP-phase, keeping $\rho=10^5$ g/cm$^3$ fixed (right panel). It is possible to see that the extra phases are (mildly) relevant only for very large densities and maximal CP-violation.

\begin{figure}[thb]
  \vspace{0.5cm}
  \centering
  \begin{tabular}{cc}
    \includegraphics[width=90mm]{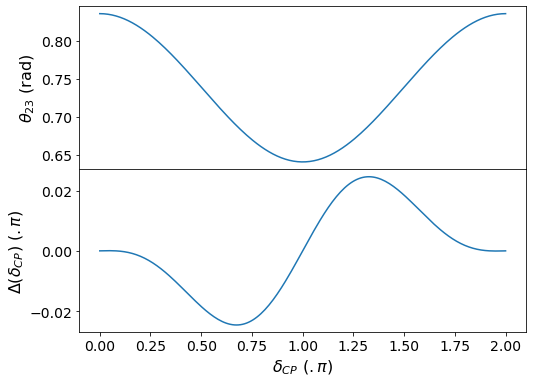} &
  \end{tabular}
    \caption{Dependence of $\tilde{\theta}_{23}$ and $\tilde{\delta}_{CP}$ on $\delta_{CP}$ for a fixed matter density ($\rho=10^5$ g/cm$^3$). In the bottom panel we show the difference $\Delta(\delta_{CP})=\tilde{\delta}_{CP}-\delta_{CP}$.}
    \label{fig:theta23}
\end{figure}

In Fig.~\ref{fig:theta23} we plot the $\tilde{\theta}_{23}$ mixing angle (top panel) and $\tilde{\delta}_{CP}$ (bottom panel) as a function of $\delta_{CP}$ for a fixed value of $\rho=10^5$ g/cm$^3$. The result is considerable: up to $\sim 10\%$ variation of the mixing angle. We note that, as highlighted in a recent work~\cite{Cherchiglia:2024bhl}, the value of $\tilde{\theta}_{23}$ for $\delta_{CP}=0$ is not the vacuum value due to the extremely high density considered.

%%%%%%%%%%%%%%%%%%%%%%%%%%%%%%%%%%%%%%%%%%%%%%%%%%%%%%
%%%%%%%%%%%%%%%%%%%%%%%%%%%%%%%%%%%%%%%%%%%%%%%%%%%%%%
\section{Analytical Considerations} \label{sec:analytical}

To have a better understanding of the need for these extra phases on the mixing matrix, it would be useful to have some analytical estimations on how to determine them. We present here one possibility, which is to analyze the jumping probability between mass eigenstates in non-adiabatic transitions, considering that such probability does not depend on the phases. Such probability can be measured, for instance, through the measurement of a day-night asymmetry in the solar neutrino flux. In Appendix A, we present to the reader some expressions relating these two quantities to the experimental results on the latest solar neutrino data regarding this
asymmetry.

The jumping probability in a non-adiabatic transition can be written as:
\[
P_{ij}=P(\nu_i\rightarrow\nu_j)=|\braket{\nu_j|\nu_i}|^2=\bigg|\sum_\alpha \tilde{U}_{\alpha j}U_{\alpha i}^*\bigg|^2
\]
where it is assumed a transition from vacuum (mixing matrix $U$) to matter (mixing matrix $\tilde{U}$). Although for the parameters we choose to work the larger jumping probability involves the second family, the analytical considerations we present here are more neat for $P_{33}$, which explicitly can be written as:
\begin{eqnarray*}
P_{33}&=&|\tilde{U}_{e3}U_{e3}^*+\tilde{U}_{\mu 3}U_{\mu 3}^*+\tilde{U}_{\tau 3}U_{\tau 3}^*|^2 \\
&=&\left|\tilde{s}_{13}s_{13}e^{i\Delta(\delta_{CP})}+\tilde{c}_{13}c_{13}\tilde{s}_{23}s_{23}e^{i\alpha}+\tilde{c}_{13}c_{13}\tilde{c}_{23}c_{23}e^{i\beta}\right|^2
\end{eqnarray*}
While the individual terms of the sum on the right-hand side show some dependence on $\delta_{CP}$, as shown in Fig.~\ref{fig:theta23}, we numerically checked that the jumping probabilities do not show such dependence. Using the constance of $P_{ij}$ when varying $\delta_{CP}$, we can write:
\[
\left|\tilde{s}_{13}s_{13}e^{i\Delta(\delta_{CP})}+\tilde{c}_{13}c_{13}\tilde{s}_{23}s_{23}e^{i\alpha}+\tilde{c}_{13}c_{13}\tilde{c}_{23}c_{23}e^{i\beta}\right|^2
=
\left|\tilde{s}_{13}s_{13}+\tilde{c}_{13}c_{13}\tilde{s}^{(0)} _{23}s_{23}+\tilde{c}_{13}c_{13}\tilde{c}_{23}^{(0)}c_{23}\right|^2
\]
where $\tilde{c}^{(0)}_{23}$ and $\tilde{s}^{(0)}_{23}$ are the values of $\tilde{c}_{23}$ and $\tilde{s}_{23}$ when $\delta_{CP}=0$. Yet not particularly intuitive, such expressions show the intrinsic relation between the new phases $\alpha$ and $\beta$ to the change of $\theta_{23}$ and $\delta_{CP}$ when neutrinos are traveling in a very dense medium.

%%%%%%%%%%%%%%%%%%%%%%%%%%%%%%%%%%%%%%%%%%%%%%%%%%%%%%%%%%%%%%%%%%%%

\section{Conclusions}\label{sec:conclusions}

The effect described in this work is negligible for the current neutrino data, since it shows up as a small effect and only at very high matter densities. However, in very dense mediums such as supernovae or the early universe, it could have an effect when performing precise numerical calculations. Therefore, besides being an important piece of the formalism on the phases of the neutrino mixing matrix, one should be reminded to include those phases when using analytical approximations for the neutrino parameters when considering high density matter effects. All numerical programs and plots mentioned in this work are available in the \href{https://github.com/GEFAN-Unicamp/Public/tree/main/2024-ExtraPhases}{GitHub repository} of our research group.

\section*{Acknowledgement}
This study was financed in part by the Coordenação de Aperfeiçoamento de Pessoal de Nível Superior - Brasil (CAPES) - Finance Code 001.

%%%%%%%%%%%%%%%%%%%%%%%%%%%%%%%%%%%%%%%%%%%%%%%%%%%%%%%%%%%%%%%%%%%
\newpage
\appendix

\section{Day-night asymmetry}

We already have an indication of a non-null transition between mass eigenstates through the measurement of a day-night asymmetry in solar neutrinos. Given the solution to the solar neutrino problem through LMA resonance, the high-energy solar neutrinos arrives at the Earth's surface roughly as $\nu_2$. When crossing from vacuum to Earth, part of this $\nu_2$ flux is converted into $\nu_1$, which results in the restart of a flavour oscillation pattern. Starting from the known expression for the electron neutrino survival probability measured during the night (assuming two neutrino families):
\[
P_{ee}^N = P_{ee}^D + \frac{1}{\cos{2\theta}}\left(1-2P_{ee}^D\right)\left(P_{2e}-\sin^2\theta\right)
\]
we can make the dependence on jumping probability explicit by replacing above the expressions:
\[
P_{2e} = P_{22}\sin^2\theta+P_{21}\cos^2\theta
\]
which leads to:
\begin{equation}
    P_{21}=\frac{P_{ee}^N-P_{ee}^D}{1-2P_{ee}^D}
    \label{eq:P21}
\end{equation}

SNO~\cite{Bellerive:2016byv} and Super-Kamiokande~\cite{Super-Kamiokande:2023jbt} are the experiments which have results on the high-energy neutrino flux day-night asymmetry. SNO fits the data with a quadratic energy dependence on the daily survival probability and a linear energy dependence on the day-night asymmetry, so it is convenient to rewrite the above expressions as:
\[
P_{21}=\left(\frac{-A_{DN}}{2+A_{DN}}\right)\left(\frac{P_{ee}^D}{1/2-P_{ee}^D}\right)
\]
where $A_{DN}=2(P_{ee}^D-P_{ee}^N)/(P_{ee}^N+P_{ee}^D)$. As a consistency check, if $P_{ee}^D<0.5$ we should expect a negative $A_{DN}$, both predictions for the LMA solution. For a $10$ MeV neutrino SNO results are:
\[
A_{DN}=0.046\pm 0.031^{+0.014}_{-0.013}
~~;~~
P_{ee}^D=0.317\pm 0.016\pm 0.009
\]
Hence, the positive asymmetry measured by SNO translates to the following limit:
\[
P_{21}< 0.02~~{(2\sigma)}
\]

While SNO results are compatible with the absence of any day-night asymmetry, thereby also with a null $P_{21}$, Super-Kamiokande data shows a mild preference for such effect. Including the neutral current events, and after some manipulation, we can write for Super-Kamiokande:
\[
P_{21}=\frac{-\bar{\Phi}(D/N)}{1+r-2\bar{\Phi}-\bar{\Phi}(D/N)}
\]
where
\[
\bar{\Phi}=\frac{1}{2}(\Phi_{day}+\Phi_{night})
~~;~~
D/N = \frac{\Phi_{day}-\Phi_{night}}{\bar{\Phi}}
\]
and $r$ is the ratio between non-electronic and electronic elastic scattering cross-section. Super-Kamiokande results are:
\begin{eqnarray*}
D/N &=&-0.0286 \pm 0.0085 (stat.) \pm 0.0032 (syst.)\\
\bar{\Phi}&=&0.445\pm 0.002 (stat.) \pm 0.008 (syst.)
\end{eqnarray*}
which leads to 
\[
P_{21}=0.042
\]

\bibliography{refs}

\begin{thebibliography}{10}

\bibitem{Pascoli:2005zb}
S.~Pascoli, S.~T. Petcov, and T.~Schwetz.
\newblock {The Absolute neutrino mass scale, neutrino mass spectrum, majorana
  CP-violation and neutrinoless double-beta decay}.
\newblock {\em Nucl. Phys. B}, 734:24--49, 2006.

\bibitem{KATRIN:2021uub}
M.~Aker et~al.
\newblock {Direct neutrino-mass measurement with sub-electronvolt sensitivity}.
\newblock {\em Nature Phys.}, 18(2):160--166, 2022.

\bibitem{Gonzalez-Garcia:2007dlo}
M.~C. Gonzalez-Garcia and Michele Maltoni.
\newblock {Phenomenology with Massive Neutrinos}.
\newblock {\em Phys. Rept.}, 460:1--129, 2008.

\bibitem{Capozzi:2021fjo}
Francesco Capozzi, Eleonora Di~Valentino, Eligio Lisi, Antonio Marrone,
  Alessandro Melchiorri, and Antonio Palazzo.
\newblock {Unfinished fabric of the three neutrino paradigm}.
\newblock {\em Phys. Rev. D}, 104(8):083031, 2021.

\bibitem{Pradhan:2024qrp}
Akhila~Kumar Pradhan, Khushboo Dixit, and S.~Uma Sankar.
\newblock {Appearance of~Majorana Phase in~Two Flavour Neutrino Oscillations
  with~Neutrino Decay}.
\newblock {\em Springer Proc. Phys.}, 304:1105--1107, 2024.

\bibitem{Benatti:2001fa}
F.~Benatti and R.~Floreanini.
\newblock {Massless neutrino oscillations}.
\newblock {\em Phys. Rev. D}, 64:085015, 2001.

\bibitem{Wolfenstein:1977ue}
L.~Wolfenstein.
\newblock {Neutrino Oscillations in Matter}.
\newblock {\em Phys. Rev. D}, 17:2369--2374, 1978.

\bibitem{Mikheyev:1985zog}
S.~P. Mikheyev and A.~Yu. Smirnov.
\newblock {Resonance Amplification of Oscillations in Matter and Spectroscopy
  of Solar Neutrinos}.
\newblock {\em Sov. J. Nucl. Phys.}, 42:913--917, 1985.

\bibitem{Cherchiglia:2024bhl}
Adriano Cherchiglia, Macello Jales, Guilherme Nogueira, Maressa~P. Sampaio, and
  Pedro~C. de~Holanda.
\newblock {Analytical Expressions for Neutrino Oscillation}.
\newblock 12 2024.

\bibitem{Bilenky:2018hbz}
Samoil Bilenky.
\newblock {\em {Introduction to the Physics of Massive and Mixed Neutrinos}},
  volume 947.
\newblock Springer, 2018.

\bibitem{Bellerive:2016byv}
A.~Bellerive, J.~R. Klein, A.~B. McDonald, A.~J. Noble, and A.~W.~P. Poon.
\newblock {The Sudbury Neutrino Observatory}.
\newblock {\em Nucl. Phys. B}, 908:30--51, 2016.

\bibitem{Super-Kamiokande:2023jbt}
K.~Abe et~al.
\newblock {Solar neutrino measurements using the full data period of
  Super-Kamiokande-IV}.
\newblock {\em Phys. Rev. D}, 109(9):092001, 2024.

\end{thebibliography}
\bibliographystyle{unsrt}

\end{document}